\documentclass[11pt]{iopart}
\usepackage{graphicx}
\usepackage{color}
\begin{document}

\title{Magnetic field mediated conductance oscillation in grapehen p-n junctions}
\author {Shu-guang Cheng}
\address{Department of Physics, Northwest University, Xi'an 710069, People's Republic of China}
\address{Shaanxi Key Laboratory for Theoretical Physics Frontiers, Xi'an 710069, People's Republic of China}
\ead{sgcheng@nwu.edu.cn}

\begin{abstract}
The electronic transport of graphene p-n junctions under perpendicular magnetic field is investigated in theory. Under low magnetic field, the transport is determined by the resonant tunneling of Landau levels and conductance versus magnetic field shows a Shubnikov de-Haas oscillation. At higher magnetic field, the p-n junction subjected to the quasi-classical regime and the formation of snake states results in periodical backscattering and transmission as magnetic field varies. The conductance oscillation pattern is mediated both by magnetic field and the carrier concentration on bipolar regions. For medium magnetic field between above two regimes, the combined contributions of resonant tunneling, snake states oscillation and Aharanov-Bohm interference induce irregular oscillation of conductance. At very high magnetic field, the system subjected to quantum Hall regime. Under disorder, the quantum tunneling at low magnetic field is slightly affected and the oscillation of snake states at higher magnetic field is suppressed. In the quantum Hall regime, the conductance is a constant as predicted by the mixture rule.
\end{abstract}
\maketitle

\section{INTRODUCTION}
Grapehne is a two dimensional material composed of hexagonal lattice carbon atoms.\cite{Graphene_1,Graphene_2} The electron band and hole band touch at the charge neural points in which the dispersion relation of the charge carrier is of linear.\cite{Dirac} The Fermi velocity of the charge carrier at the charge neutral point is independent of the energy and acting as massless relativity particles. Extremely high charge carrier mobility in graphene is realized in graphene-hexagonal boron nitride heterostructures.\cite{High_m1,High_m2} Interesting phenomena are found in monolayer graphene, such as anomalous Hall effect\cite{Graphene_Hall1,Graphene_Hall2} and Klein tunneling.\cite{Klein, Klein1, Klein2} The charge carrier in graphene can be manipulated through doping or gate voltage to be electron-like or hole-like.\cite{review} So in a graphene nanoribbon, a p-n junction is realized by applying negative/positive gate voltage in two sides.\cite{Veselago_lens,Veselago_lens1, Graphene_pn1,Graphene_pn2,Graphene_pn3,Graphene_pn8,Graphene_pn6,Graphene_pn4,Graphene_pn5,Graphene_pn7} Electrons can transmit through a p-n-p junction interface with comparatively large transmission coefficient by Klein tunneling.\cite{Klein2,pnp-1,pnp-2} Due to the optical analogy of Dirac fermion, a Veselago lens is realized. \cite{Veselago_lens,Veselago_lens1} Through electrostatic gating, both positive and negative refraction across the graphene are found, in good agreement with the Snell's law prediction.\cite{Graphene_pn4}

Under strong magnetic field, the propagation directions of the edge states at bipolar regions are opposite to each other.\cite{copropa} Thus at p-n junction interface, the charge carriers are co-propagating. In clean devices, the Klein tunneling induces perfect transmission when the incident charge carriers are normal to the interface.\cite{Graphene_pn6} Driven by Lorentz force, the cyclotron orbit of charge carrier is normal to the p-n interface. In this process, the charge carrier experiences between electron-like and hole-like through Klein tunneling and forms snake states.\cite{snake1,snake3,snake2,snake5,snake4} The snake states also exist in two dimensional electron gas when the magnetic field is nonuniform in space.\cite{snake_1} For a unipolar system, such as a folded graphene or a carbon nanotube, a uniform magnetic field can induce snake states at the lateral side in which the effective magnetic field is zero.\cite{snake_2,snake_3} For a graphene p-n junction in the presence of strong magnetic field and disorder, the charge carriers are redistributed at the interface, leading to the conductance mixture rule of $G=|\nu_1||\nu_2/|(|\nu_1|+|\nu_2|)G_0$ with $\nu_{1/2}$ the filling factor of both sides and $G_0=e^2/h$ the quantum conductance.\cite{Graphene_pn1, Graphene_pn2,Graphene_pn3} Due to the edge scattering \cite{edge_scatter} and disorder induced intervalley and intravalley scattering,\cite{intervalley} the mixture rule holds true for all transmission coefficients between Landau levels in p and n regions.\cite{Daining} The charge carrier mixture rule applies both to monolayer graphene of chiral edges and bilayer graphene under Anderson disorder and electron-hole puddles.\cite{pn-1,pn-2}

Very recently in a work by Ke. et al,\cite{network} the resonant tunneling of quantum Hall states is studied in a bilayer graphene p-n-p network. The transmission oscillation versus magnetic field, found in comparatively low magnetic field, comes from the resonant tunneling of Landau levels in bipolar regions, mediated by magnetic field and the gate voltages. Another type of conductance oscillation found in graphene p-n junction is contributed by snake states.\cite{snake3} When the cyclotron diameter of charge carriers is changed by magnetic field or gate voltage, at the end route, the charge carriers can either be backscattered or transmit, determined by cyclotron diameter and the p-n junction width.\cite{snake2} Consequently, conductance oscillation happens by varying magnetic field or sample width or charge carrier concentration at bipolar regions. In an experiment work, the Aharanov-Bohm (AB) interference is reported in a graphene p-n junction under magnetic field.\cite{Graphene_pn5} It is claimed that the interference pattern is determined by the magnetic field magnitude and the charge carrier concentrations, which combine alter the effect area enclosed by the edge channels. The conductance at high magnetic field deviates from the result given by mixture rule in the presence of disorder. Thus in the graphene p-n junction, we wonder how the three mentioned transmission oscillations function as the magnetic field strength varies. Do they exist independently or act simultaneously? Besides, how the oscillation is affected by the carriers concentration, sample width or different chiral graphene ribbons? Also, in the quantum Hall regime how the conductance evolves from oscillation in a clean model to a constant determined by mixture rule in the presence of disorder is interesting as well.

In this work, a graphene nanoribbon p-n junction subjected to a perpendicular magnetic field is investigated by using the tight-binding model and Green's function method. At low magnetic field, the graphene p-n junction is in quantum tunneling regime and the transmission happens on one edge of p-n interface. The conductance decreases as the magnetic field is enhanced and shows a Shubnikov de-Haas type oscillations. The period of the oscillation is large (small) when the charge carrier concentrations in the bipolar regions are small (large). At higher magnetic field (quasi-classical regime), when the cyclotron radius of charge carrier is smaller than the sample width, the snake states are formed. The conductance versus the magnetic field shows an oscillatory pattern. At stronger magnetic field or lower charge concentration the period shrinks, indicating the decrease of the cyclotron radius. Between quantum tunneling regime and quasi-classical regime, the conductance shows an irregular oscillation. It is induced by three mechanisms: quantum tunneling, snake states and AB effect. The effects of sample width, charge concentration and graphene ribbon of different chirality are investigates. In the presence of disorder, the conductance oscillation in the quantum tunneling region is slightly affected. In the quasi-classical regime, however, the snake states induced conductance oscillation is suppressed. In the quantum Hall regime, the conductance is a constant determined by the mixture rule.

The paper is organized as follows. In section \ref{Modol}, the Hamiltonian and the method are introduced. Main results are given in section \ref{results}, including the oscillation of the graphene p-n junction conductance in different regimes and the dependence relations with the magnetic field, the gate voltage and sample size. The situations for devices of different edges and the presence of Anderson disorder are also discussed. A brief discussion is given in section \ref{conclusion}.
\section{MODEL AND METHOD}\label{Modol}
First we take a zigzag edged graphene p-n junction as an example and the chiral gaphene ribbons are discussed in later discussion. The model is displayed in Fig. \ref{Fig1} (a): two infinite terminals are connected by a central region. The charge concentrations in two terminals are controlled by gate voltages: $V_L$ in the left terminal and $V_R$ in the right terminal. In the central region the potential changes from $V_L$ to $V_R$ depending on specific cases [see Fig. \ref{Fig1} (b)]. The Hamiltonian in the tight-binding representation is written as:\cite{Graphene_pn3,H1,H2}
\begin{eqnarray}
H=\sum_{i}\epsilon_i a_i^\dagger a_i-\sum_{<ij>}(te^{i\phi}a_i^\dagger a_j+te^{-i\phi}a_j^\dagger a_i)
\label{EQ1}
\end{eqnarray}
The first term is the on-site energy and the second term means the nearest hopping with energy $t$. Under a perpendicular magnetic field, the phase induced by vector potential $\mathbf{A}$ is $\phi_{\mathbf{i}\mathbf{j}}=\frac{2\pi e}{h}\int_{\mathbf{i}}^{\mathbf{j}} \mathbf{A}\cdot d\mathbf{l}$. Under a gauge transformation we set the vector potential circling a single hexagonal lattice is $\phi_0=\frac{\pi e}{h}\oint \mathbf{A}\cdot d\mathbf{l}$. So the actual magnetic induction strength $B$ is connected with $\phi_0$ through a relation $BS=\oint \mathbf{A}\cdot d\mathbf{l}=\frac{h\phi_0}{\pi e}$ with $S$ the area of a single hexagonal unit. For instance, when $\phi_0=0.001$, the actual magnetic field is $B=\frac{2h\phi_0}{3\sqrt{3}\pi e a^2}\simeq25.3 T$ ($a=0.142nm$ the C-C bond length of graphene), which is hard to access in experiment. According to the scaling approach,\cite{scalling1,scalling2} when the sample size becomes large, weaker magnetic field is required to obtain the reliable results.

Using the Green's function method, the transmission coefficient through a p-n junction reads: \cite{LB_form}
\begin{eqnarray}
T(E)=\mathbf{Tr}(\mathbf{\Gamma}_L\mathbf{G}^r\mathbf{\Gamma}_R\mathbf{G}^a)
\label{EQ2}
\end{eqnarray}
with $\mathbf{\Gamma}_{L/R}$ the line-width function of terminal $L/R$ and $\mathbf{G}^r/\mathbf{G}^a$ the retarded/advanced Green's function. $\mathbf{G}^r$ is calculated from the relation $\mathbf{G}^r=(E\mathbf{I}-\mathbf{H}_c-\mathbf{\Sigma}^r_L-\mathbf{\Sigma}^r_R)^{-1}$ with $\mathbf{H}_c$ the Hamiltonian of the central region and $\mathbf{\Sigma}^r_{L/R}$ the retarded self-energy of left/right terminal. The line-width function is determined by $\mathbf{\Gamma}_{L/R}=i(\mathbf{\Sigma}^r_{L/R}-\mathbf{\Sigma}^a_{L/R})$. Finally $\mathbf{\Sigma}^r_{L/R}$ can be calculated numerically by different ways.\cite{sur1,sur2} At zero temperature, the linear conductance $G(E)$ is deduced from the Landauer-B\"{u}ttiker formula\cite{LB_form} that $G(E)=\frac{e^2}{h}T(E)$ with no consideration of spin degeneracy.

In the numerical calculation we set the Fermi level $E=0$. In the left (right) terminal $\epsilon_i$ in equation \ref{EQ1} is $V_L$ ($V_R$) in unit of $t$. In the central region, $\epsilon_i$ varies from $V_L$ to $V_R$, depending on the specific case. When Anderson disorder is considered, in the central region $\epsilon_i$ at each site is added with a random potential subjected to a uniform distribution $[-W/2, W/2]$ with $W$ the disorder strength.\cite{dis1} For each $W$ the conductance is average to $500$ configurations.
\section{Numerical results}\label{results}
We investigate the oscillation characters of graphene p-n junction and explain the mechanisms related in subsection \ref{sub1}. The behavior of conductance oscillation affected by gate voltage, sample length and sample width are detailed in subsection \ref{sub2}. The situations for chiral graphene ribbons and the presence of Anderson disorder are investigated in subsection \ref{sub3}.

\subsection{The characters of the conductance oscillations}\label{sub1}

\begin{figure}
\begin{center}
\includegraphics[width=10cm, viewport=147 3 565 510, clip]{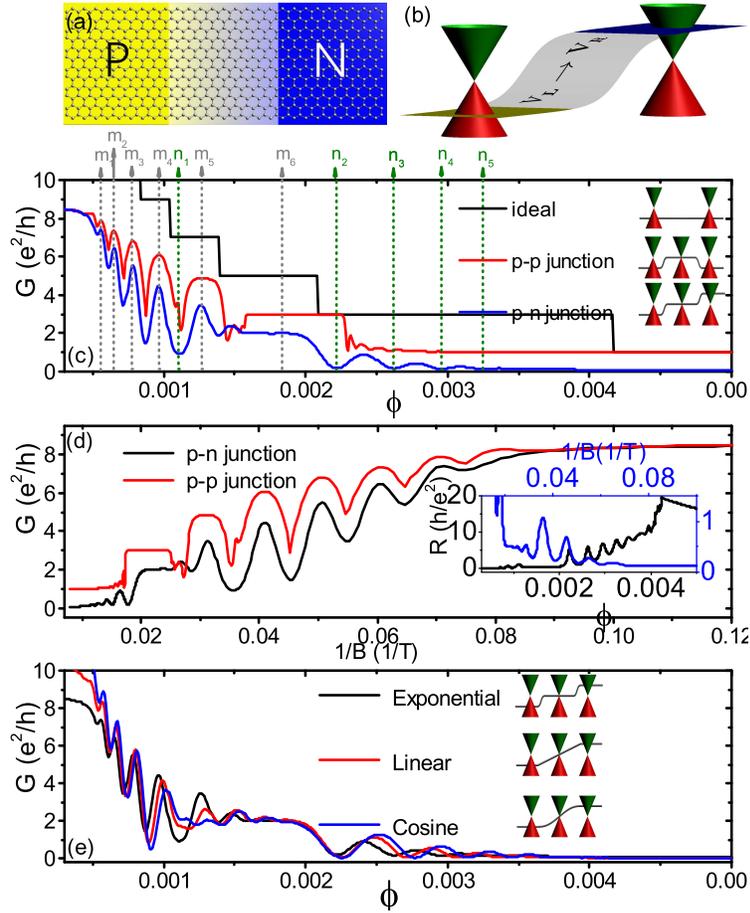}
\caption{(color online) (a) The schematic of graphene nanoribbon p-n junction. The width of ribbon is $N=6$ and the length of the central region is $M=9$. (b) The schematic of band structure and potential drop in space. In the left (right) terminal the carrier is of hole-like (electron-like) and the potential drops exponentially in the central region from $V_L$ to $V_R$. (c) Magnetic field dependence of conductance for three cases: ideal ribbon, p-p junction and p-n junction. The insets follow the legend are the band structures of the corresponding p-n junction and potential drop (grey lines) in space. (d) Magnetic field ($1/B$) dependence of conductance for p-n and p-p junctions. The inset shows $R$ vs. $\phi$ and $1/B$ of a same p-n junction in (d). (e) Magnetic field dependence of conductance for three types of potential variation in the p-n interface: exponential, linear and cosine. In (c-e) the parameters are $N=150$, $M=20$ and $V_L=-V_R=0.12t$.
}\label{Fig1}
\end{center}
\end{figure}

The magnetic field dependence of conductance for an ideal graphene nanoribbon, a p-p junction and a p-n junction are shown in Fig. \ref{Fig1} (c). The parameters are $N=150$ and $M=20$. The conductance for an ideal model shows quantized values and it tells the filling factor for each terminal of a p-n or p-n junction. Since we have $|V_L|=|V_R|$ and the system is invariant under electron hole inversion. In experiment, the charge concentration in the interface of p-n junction does not change abruptly from p-type to n-type. Thus in our model, the electric potential in the central region of p-n junction decreases exponential from $V_L$ to zero and undergoing a deplete area and again decreases exponential from zero to $V_R$ [see the bottom inset in \ref{Fig1} (c)].

\begin{figure}
\begin{center}
\includegraphics[width=15cm, viewport=30 44 690 510, clip]{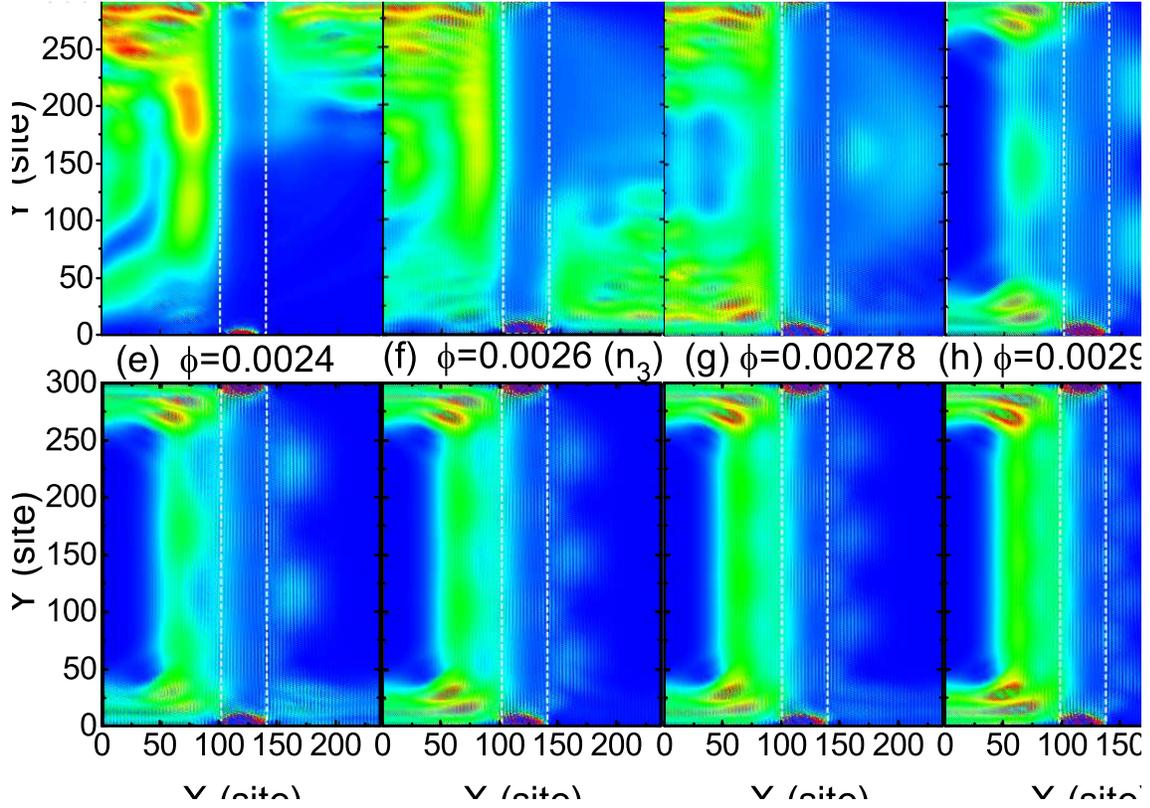}
\caption{(color online) Spacial distribution of local density of states in graphene p-p junction (a) and p-n junction (b-h) at different magnetic field. The corresponding $\phi$ value is displayed at the top of each figure and the index $n_i$ is from Fig.\ref{Fig1} (c). (b) The potential varying area is highlighted with white rectangles. The other parameters are the same to a p-n junction in Fig.\ref{Fig1} (b).}\label{Fig2}
\end{center}
\end{figure}

At small $\phi$, $G$ decreases and shows a Shubnikov de-Haas oscillation in p-n junction. At low magnetic field, the Landau levels start to form and edges states are not formed yet. The transmission is contributed by quantum tunneling of Landau levels.\cite{network} The resonant peaks are indexed in Fig.\ref{Fig1} (c) as $m_1-m_6$. When the Landau levels at both terminals are around the Fermi level, resonance peak appears. In contrast, when in both terminals the Fermi level is located between two neighbor Landau levels, off-resonance happens. For comparison, the result for a p-p junction is shown as well because in such a case, only resonant tunneling of Landau levels is allowed. The electric potential in the central region of p-p junction decreases from $V_L=V_R=0.12t$ at the edges to zero in the center. Similar oscillation pattern is seen in low magnetic field with the same location of peaks and dips. With the help of local density of states calculation [$\rho(\mathbf{r},E)= (\mathbf{G}^r\mathbf{\Gamma}_L\mathbf{G}^a)_{\mathbf{r}}/2\pi$\cite{DOS} with $\mathbf{G}^{r/a}$ the Green's function and $\mathbf{\Gamma}_L$ the line-width function in equation (\ref{EQ2})], the transmission in space is more clear. In a p-p junction, the cyclotron directions of the charge carriers at both terminals are the same. In Fig. \ref{Fig2}(a), left coming charge carriers on the upper edge are either backscattered to the bottom edge of the left terminal, or transmit to the right terminal, on the upper edge. In p-n junction, the charge carriers are transmitted into the bottom edge of the right terminal. They indicate that at low $\phi$ the transmission is determined by quantum tunneling of Landau levels. It can also be clearly seen from Fig. \ref{Fig1} (d) in which $G$ vs. $1/B$ shows peaks of equal distance. In inset of Fig. \ref{Fig1} (d), $R$ vs. $1/B$ relation is shown with the blue line. Resonant peaks of equal spacing in $1/B$ axis are seen which is reported in a recent experiment.\cite{network}

\begin{figure}
\begin{center}
\includegraphics[width=15cm, viewport=4 173 740 526, clip]{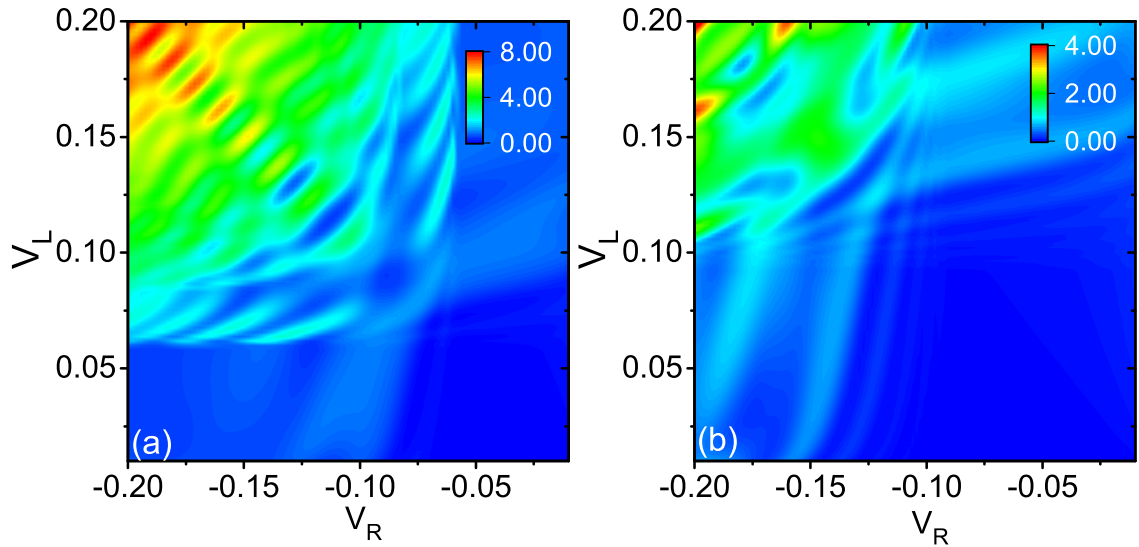}
\caption{(color online) $V_L$ and $V_R$ dependence of p-n junction conductance in the quantum tunneling regime (a) and quasi-classical regime (b). The other parameters are $N=150$, $M=20$ and $\phi=0.001$ in (a) and $\phi=0.0025$ in (b).
}\label{Fig3}
\end{center}
\end{figure}

To further verify the assumption of quantum tunneling induced conductance oscillation, the $V_L$ and $V_R$ dependence of $G$ in two regimes (quantum tunneling and quasi-classical regimes) are displayed in Fig.\ref{Fig3}. In Fig.\ref{Fig3} (a), a clear chessboard oscillation patterns are seen as $V_L$ and $V_R$ changes. It demonstrates that the transmission is in the quantum tunneling regime. When $V_L$ and $V_R$ move in the same direction (both increase or decrease with magnitude) or opposite direction (one increases and another decreases with magnitude), the conductance shows a maximum when the Landau levels in p and n region cross the Fermi level ($E=0$). At small values of $|V_L|$ and $|V_R|$, no obvious oscillation is seen because there is no Landau level cross the Fermi level.

In the interval $\phi\in[0.0012, 0.0018]$, the oscillation of $G$ for a p-n junction shows irregular pattern, contributed by the coexistence of resonant tunneling, snake states and AB interference (detailed in later discussions). In this transition regime the cyclotron radius of the charge carriers is of the same magnitude with sample width. This interval lies between $n_1$ and $n_2$ in Fig.\ref{Fig1} (c), ie, when the cyclotron radius of the charge carrier is smaller than the width of the ribbon but larger than half of the width of the ribbon. The conductance for p-p junction shows a quantized conductance of $G=3G_0$. This quantized value, usually smaller than the filling factor, means that only limited number of states can transmit through the potential barrier. In fact the spacial distributions of different Landau level are different as well.\cite{view_LL} In the p-p junction the Landau level with wave function expanded wider in space is more easily backscattered and makes no contribution to the transmission. In contrast, the Landau level with narrower expansion can easily tunnel through the central region and make contribution to the conductance. This result is checked by the scattering matrix method and not shown.

Start from $\phi=0.002$, the system subjected to quasi-classical regime and snake states are formed in the p-n interface. $G$ vs. $\phi$ shows an oscillation and both the period and the magnitude of the oscillation decrease as $\phi$ is enhanced. The dip value is symbolled by $n_i$ in Fig.\ref{Fig1} (c) with $i$ the number of cyclotron orbit in the right terminal. The spacial distributions of local density of states for snake states at different $\phi$ are shown in Fig.\ref{Fig2}(c-h). In Fig.\ref{Fig2}(c), (d), (f) and (h), there are $1$, $2$, $3$, and $4$ whole cyclotron orbits in the right terminal edge, respectively. The conductance shows minimal values. At the end route of snake state, charge carriers are scattered to the left terminal and the density of states from the left terminal is nearly zero at the bottom of right terminal. In (e) and (g), there is an integer plus half number of cyclotron orbits in the right terminal edge and the conductance shows peaks. Here the final route of charge carriers are scattered to the right terminal. The absence of quantum tunneling can also be seen from the contour map of $G(V_L, V_R)$ in Fig.\ref{Fig3}(b). In this regime there are few Landau levels below the Fermi level [see Fig.\ref{Fig1} (c)] and conductance is mainly determined by snake states.

As $\phi$ becomes very large, the system is under quantum Hall regime. $G$ is nearly zero (deviated from the mixture rule) because in a clean zigzag edged graphene p-n junction, the edge states are all backscattered due to the antiparallel isospins in both edges. The resistance $R$ vs. $\phi$ relation is shown in inset of Fig.\ref{Fig1} (d) (curve in black). The curve is very similar to the experimental result obtained in an ultra clean graphene p-n junction.\cite{Graphene_pn5} In the experiment there are tiny resistance oscillation at low magnetic field [see Ref \cite{Graphene_pn5} FIG. 2 and Ref \cite{snake3} Figure. 4 (a)] which should be contributed by resonant tunneling of Landau levels.

To find out whether the $G-\phi$ oscillatory behaviors are affected by the potential variation in the central region, we focus three types of potential changes in the central region as shown in insets of Fig. \ref{Fig1} (e): exponential, linear and cosine. At very small magnetic field, $G$ for cosine-type (exponential) potential is the largest (smallest). It is quite reasonable because for a given length of central region, exponential potential changes most abruptly and departs the bipolar regions best. On the other hand, cosine potential is the smoothest one and bipolar regions are separated inadequately. When $\phi$ is small, the peaks for three cases are almost the same. However, a small departure comes that the distance between nearby peaks is the largest for cosine potential and are the smallest for exponential potential. It can be explained by the edge potential modified Landau level spacing. In the transition regime the curves have no obvious difference. In the quasi-classical regime, all $G-\phi$ curves show oscillating and damping behaviors. The minimum position of $G$ do not coincide with each other and it could be contributed by the deformation of cyclotron orbits due to the density variation in the central region.\cite{snake3} In the resting discussion we choose exponential potential as an example.

So far, $V_L=-V_R$ is assumed in above discussion. If $|V_L|\neq|V_R|$, two set of resonant peaks in quantum tunneling regime and snake states oscillation in quasi-classical regime can be seen (results not shown) and the explanations are the same. In the follow discussion, if we adopt $V_L=-V_R$, we use $V_0=V_L=-V_R$ for simplicity.
\subsection{Gate voltage and sample size dependence of conductance oscillation}\label{sub2}
\begin{figure}
\begin{center}
\includegraphics[width=10cm, viewport=100 45 537 575, clip]{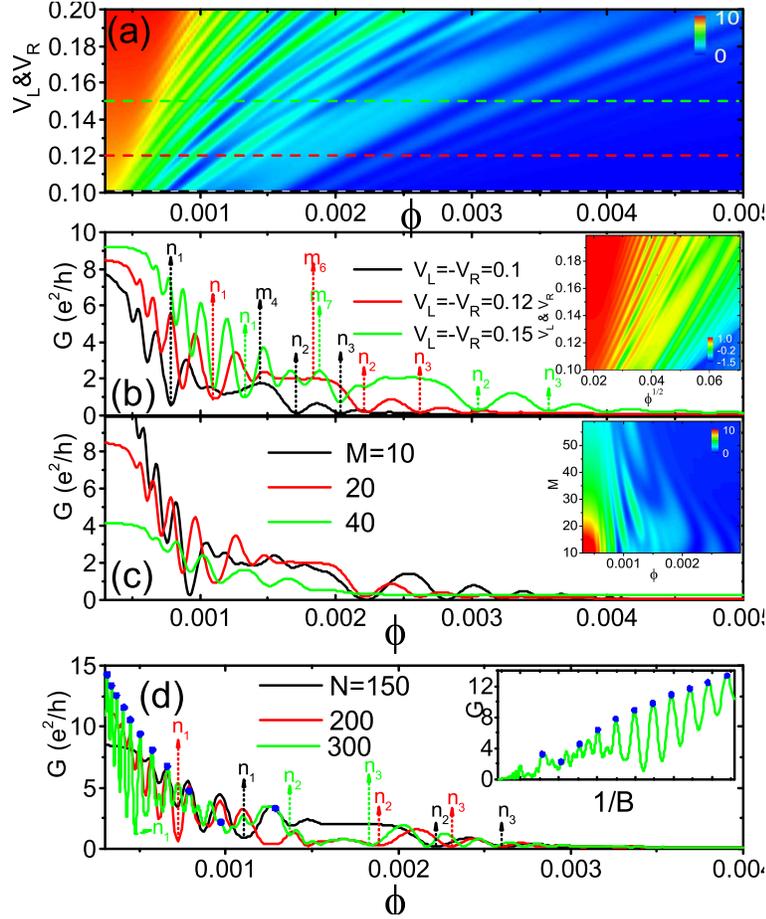}
\caption{(color online) (a) Contour plot of $G$ of graphene p-n junction on $V_0$ and $\phi$. Here we use the parameters $V_0=0.12t$, $N=150$ and $M=20$. To show clearly the oscillation of $G$, $\log(G/G_0)$ versus $V_0$ and $\sqrt{\phi}$ is shown in in set of (b). The cross-sections of (a) are shown in (b) with the final Landau level in quantum tunneling regime and first three $G$ dips in quasi-classical regime indexed as $m_i$ and $n_j$, respectively. $G$ vs. $\phi$ for different $V_0$ (b), $M$ (c) and $N$ (d). The contour plot of $G(\phi, M)$ is shown in inset of (c). In (c) $N=150$ and in (d) $M=20$.
}\label{Fig4}
\end{center}
\end{figure}

In above discussion, we learned that the behavior of graphene p-n junction conductance in the magnetic field could be classified into four regimes: quantum tunneling, transition regime, quasi-classical regime and quantum Hall regime. In this subsection we investigate how the four regimes are determined by related parameters. In Fig.\ref{Fig4} (a), the conductance dependence on $V_0$ and $\phi$ are displayed. At low magnetic field, when $V_0$ is small, quantum resonance happens with few resonance peaks. As $V_0$ increase, quantum resonance happens in wider magnetic field interval with more Landau levels involved. At higher magnetic field, the conductance oscillation induced by snake states happens. For larger charge concentration, the conductance oscillation happens at higher magnetic field. When $V_0$ is large, stronger magnetic field is asked to guarantee a same cyclotron radius of charge carriers. Meanwhile, for large $V_0$, the transition regime moves to a interval with stronger magnetic field. The $G(V_0, \phi)$ relation is reconstructed as $\log G/{G_0}$ versus $V_0$ and $ \sqrt\phi$ in Fig.\ref{Fig4} (b) inset for better view. At large $V_0$ and $\phi$, the oscillation pattern in quantum tunneling regime and quasi-classical regime is roughly proportional to $\propto V_g/\sqrt{\phi}$. The conductance oscillation pattern is deviated from the parabolic pattern in Ref \cite{snake3}. It may come from the different relation between gate voltage and charge concentration.

The $V_{0}$ dependence of $G$ can be seen more clearly in Fig.\ref{Fig4} (b) [correspond to dashed lines in Fig.\ref{Fig4} (a)]. In general, for large $V_0$, the curve is shifted rightwards. For each $V_0$, the $\phi$ values correspond to the final Landau level in quantum tunneling regime and the first three $G$ dips in quasi-classical regime is indexed by arrows and letters of the same color. When $V_{0}$ increases, all indexes move to large $\phi$ values. Besides, a new character comes that for larger $V_{0}$ there are more irregular oscillation appears in the transient regime (between $n_1$ and $n_2$). Since for larger $V_0$, the transient regime is expanded, we attribute the new appeared complex conductance oscillation to AB interference. In Fig.\ref{Fig2}, at the bottom end of the p-n interface there is scattering center, the carriers can flow back to the left terminal. Similarly, the charge carriers can also be transmitted by the scattering center from left terminal to right terminal via the bottom of the p-n interface and consequently, an enclosed AB interferometer forms. So in the transient regime, except for the quantum tunneling, both snake states and AB interference are involved. In Ref \cite{snake2}, by changing the ribbon width, the charge carriers flow into different terminals changes periodically. However in this process, the area enclosed by the AB interferometer is increased simultaneously, the role of AB interference cannot be eliminated. If we keep the ribbon width unchanged and vary the central region length $N$, the transmission of snake states might be suppressed, but the effect of AB interference will be amplified. To demonstrate this assumption, a two dimensional map of $G$ versus $\phi$ and $M$ is shown in Fig.\ref{Fig4} (c). At small given $\phi$, when $M$ becomes large, $G$ decreases due to the existence of potential barrier in the central region. There is no oscillation pattern because the quantum tunneling happens mainly on the upper side of the p-n interface. For large $\phi$, no obvious oscillation of $G$ is found because the transmission at bottom of the p-n interface is small, especially for large $M$. In the transient regime, however, clear oscillation of $G$ appears. The results are similar to what observed in experiment:\cite{Graphene_pn5} for large carrier concentration conductance oscillation happen at stronger magnetic field. But the explanations are different. In fact in Ref \cite{Graphene_pn5}, the possibility of snake states induced conductance oscillation could not be eliminated.

How the oscillation of $G$ is affected by the sample size? In Fig.\ref{Fig4}(c-d) the results are displayed. In Fig.\ref{Fig4} (c), $G$ vs. $\phi$ for different $M$ is shown. At very low magnetic field, longer central region results in smaller transmission coefficients due to the prohibition of central potential barrier. In the quasi-classical regime, the number of $G$ peaks for different $M$ is the same. When $M$ is smaller, the resonance oscillation is more obvious (larger peak values and smaller dip values). In the transient and quasi-classical regime, the AB interference and snake states begin to function. Distinct differences of $G$ oscillations appear: i) the magnitude of the oscillation is strong for small $M$ and disappears for large $M$ (e.g. $M=40$); ii) at a same $\phi$, the period of oscillation is smaller for larger $M$; iii) more irregular conductance oscillations appear when $M$ is small. Those items are reasonable. For large $M$, the area enclosed by edges states becomes large and the $\phi$ difference needed for an extra quantum flux is small. Also the enlarged distance between bipolar regions decreases the transmission coefficient and hence the magnitude of the oscillation is damped. When $M$ is small, AB interference dominate and there are more resonant peaks.

The results for different ribbon width ($N$) are displayed in Fig.\ref{Fig4}(d). When $N$ is large, there are more Landau levels below the Fermi level and more resonant peaks appears at low magnetic field [see the blue dots marked in the curve $N=300$ and inset of Fig.\ref{Fig4} (d)]. In Fig.\ref{Fig4} (d) inset, the $G$ vs $1/B$ relation is displayed. Blue dots are used to symbol the resonance peaks of equal spacing in $1/B$ axis. At small $B$, the $G$ peaks of equal distance are clear. The indexes for the first three integer number of cyclotron orbits are marked as $n_1, n_2$ and $n_3$. For wider sample, the same cyclotron radius is formed at lower magnetic field. In the transient regime, $G$ for $N=150$ is large because the ribbon includes one and a half cyclotron orbit. While for $N=200$ and $N=300$, $G$ is small which should be attributed to the destructive interference of AB effect. In the quasi-classical regime, the amplitude of the $G$ oscillation at the same $\phi$ for different $N$ is almost the same because of the roughly the same transmission coefficient through a central region of the same length. Here one may suppose that the period of oscillation for different $M$ should be the same since there are only couples of more cyclotron orbits at the interface (e.g. $n_2$ for $N=200$ is close to $n_3$ for $N=300$ and $n_2$ for $N=150$ is close to $n_3$ for $N=200$). However, the period of the $G$ oscillation shrinks obviously for large $N$. It should be attributed to extra phase accounted by AB effect due to the enlarged area enclosed by AB interferometer.

\subsection{Chiral graphene p-n junctions and effects of disorder}\label{sub3}
\begin{figure}
\begin{center}
\includegraphics[width=10cm, viewport=106 106 693 515, clip]{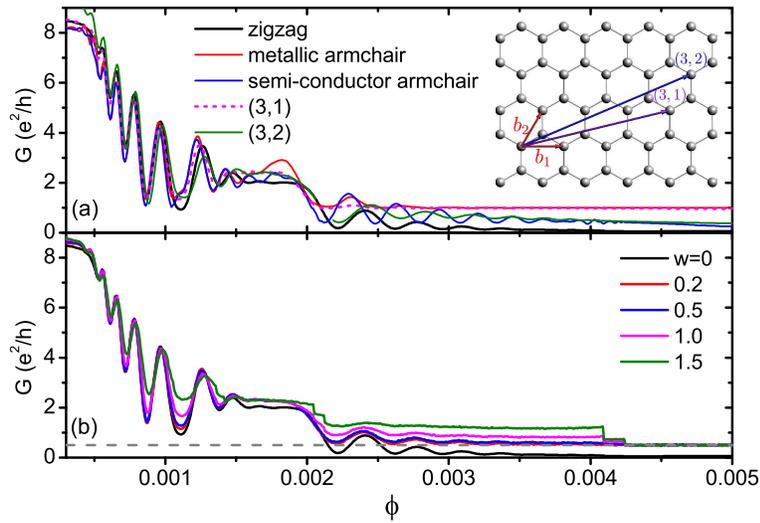}
\caption{(color online) $G-\phi$ relation in graphene p-n junction of different edges (a) and different disorder strength (b). In (a) the sizes of all five devices are selected that the actual width and central region length are the same. The inset in (a) shows the hexagonal lattice the basis vectors and chiral edges of $(3,1)$ and $(3,2)$. In (b), the parameters are $N=150$, $M=20$ and the dashed line indicates $0.5G_0$. In both (a) and (b), $V_0=0.12t$.
}\label{Fig5}
\end{center}
\end{figure}

In above discussion, only the zigzag edged clean graphene p-n junctions are investigated. In the following we discussion situations when the graphene p-n junctions are of chiral and armchair edged. In the inset of Fig.\ref{Fig5} (a), a graphene hexagonal lattice and the basis vectors of the triangular Bravais lattice are shown as $\mathbf{b}_1$ and $\mathbf{b}_2$. The orientation of the graphene ribbon is expressed by the combination of $\mathbf{b}_1$ and $\mathbf{b}_2$.\cite{chiral1,chiral2} In Fig.\ref{Fig5} two types of chiral graphene ribbon are investigated, including $(3, 1)$ and $(3,2)$ type edges. For the case of armchair edges graphene, we adopt two types, a metallic one and a semiconductor one. The result for zigzag edged grpahene p-n junction is also adopted for comparison. Here for all five types of graphene p-n junction, the actual widths are the same and the magnetic field parameter $\phi$ is scaled due to the same strength of the magnetic field.

All the curves are similar in the quantum tunneling regime: including the resonance peaks' magnitudes and positions. In the transient regime, irregular but small oscillation appears but the curves are roughly of the same character. One also notes that the magnetic field intervals of the transient regime for five ribbons are also the same. In the quasi-classical regime, three types of ribbon (zigzag edges, metallic armchair edged and chiral $(3,2)$ graphene ribbon) show a similar oscillatory and damping behavior with nearly equal period. However, for the cases of the semiconductor armchair edged and chiral $(3,1)$ graphene ribbon, $G$ is close to $G_0$ and the oscillation is nearly invisible. The perfect transmission happens because the valley-isospins at the two edges of the left terminal equals to $\pi$, hence the backscattering is forbidden.\cite{Graphene_pn8,edge_scatter} In the quantum Hall regime, $G$ for zigzag edge model decays to zero and $G$ for metallic armchair edged and chiral $(3,2)$ graphene ribbon is large because of the partition rule decided by isospins difference at both edges of the ribbon.\cite{isospin1,isospin2}

The effect of Anderson disorder on the p-n junction is investigated in Fig.\ref{Fig5} (b). In the quantum tunneling regime, the resonance oscillation is barely affected by weak disorder. When the disorder is strong (e.g. $W=1t$), the peaks' values are unaffected and the dips' values are enhanced. As a result the amplitude of the resonant oscillation decreases. In this regime, the movement of the charge carriers is driven by Lorentz force. For the case of resonance tunneling (peaks in curve), charge carriers in the central region are scattered but the total transmission is robust against disorder. In contrast, for the case of off-resonance tunneling (dips in curve), Anderson disorder provide additional opportunity for charge carriers scattered from the left terminal by way of nearby Landau levels, to the right terminal. In the transient regime, $G$ is only slightly enhanced and the irregular oscillation disappears because the AB interference is damaged by disorder induced scattering. Besides, the whole transient regime is not affected by disorder in the $\phi$ axis. In the quasi-classical regime, the snake states induced conductance oscillation is strongly affected by disorder. For weak disorder strength (e.g. $W=0.2t$ and $0.5t$), $G$ increases and the oscillation becomes obscure. The suppression effect is much stronger at larger $\phi$ in which $G\sim 0.5G_0$, indicating the effect of mixture rule.\cite{Graphene_pn1} In this case only one Landau level is involved in the total mixture process. For stronger disorder (e.g. $W=1.0t$ and $1.5t$), a second Landau level (filling factor $\nu=3$, see Fig.\ref{Fig1} (c) for ideal case) takes part in the mixture and the conductance is larger than $0.5G_0$. However, the mixture of the second Landau level is not complete thus $G$ is less than $1.5G_0$.\cite{Graphene_pn7}
When $\phi>0.0042$, there is only one Landau in each terminal [see Fig.\ref{Fig1} (c)], the complete mixture happens: $G=0.5G_0$ for $W$ increases from $0.2$ to $1.5$.

\section{Discussion and Conclusion}\label{conclusion}
The conductance oscillation of graphene p-n junctions under magnetic field is investigated in theory. Depending on the magnitude of magnetic field, the behaviors of $G$ are classified into four categories: the quantum tunneling regime, the transient regime, the quasi-classical regime and the quantum Hall regime. The effect of potential profile at p-n interface, the charge carrier concentration in p or n region, the sample size of the model and ribbons of chiral edges on the conductance oscillation are discussed.

In the quantum tunneling regime, Shubnikov de-Haas conductance resonance happens due to the transmission between Landau levels. It is observed in a bilayer graphene p-n networks very recently.\cite{network} In the transient regime, the combination of quantum tunneling, snake states and AB effect result in irregular oscillation of conductance. In the quasi-classical regime, the conductance oscillation is determined by snake states and is also affected by the AB interference. The snake states in a ballistic suspended graphene p-n junction was found through transport measurement at low magnetic field.\cite{snake3} In another work by Morikawa {\it et al}, the AB interference is reported.\cite{Graphene_pn5} The presence of snake states and AB interference is due to the high quality of graphene p-n junction. The characters of conductance oscillations depending on magnetic field and gate voltage are in good agreement with our numerical results in the clean limit. In the quantum Hall regime, the transmission is nearly zero for a clean device of zigzag edge and the perfect transmission happens when the device is of metallic armchair and some chiral edges. The different edge dependent transmission is determined by the isospins at both edges. In experiment the isospin is modulated by the gate voltages in a graphene p-n junction of rough edges and the conductance oscillates.\cite{Graphene_pn8} In the presence of disorder, the resonant tunneling induced conductance oscillation is slightly suppressed and snake states induced conductance oscillation is smeared out. When both magnetic field and disorder are strong, at the p-n junction interface, the mixture of quantum Hall states is acting: the lowest Landau level is fully mixed and higher Landau levels are mixed incompletely. In an early experimental work, the partition of quantum Hall states in graphene p-n junction with disorder is reported.\cite{Graphene_pn1} The different mixture characters of Landau levels are investigated in a recent experiment.\cite{Graphene_pn7}
\section*{ACKNOWLEDGMENTS}
This work was supported by NSFC under Grants No. 11674264, 51572219 and 11447030.

\end{document}